\documentclass[prb,twocolumn,superscriptaddress,nofootinbib,
    nobalancelastpage]{revtex4}

\usepackage{amssymb,amsmath,bm,graphicx,psfrag,color}

\newcommand{\poly}{\mathrm{poly}}
\newcommand{\ket}[1]{|#1\rangle}
\newcommand{\bra}[1]{\langle#1|}
\newcommand{\tr}{\mathrm{tr}}
\newcommand{\dd}{\mathrm{d}}
\newcommand{\QMA}{\textsf{QMA}}

\newcommand{\NP}{\textsf{NP}}

\newcommand{\V}{\mathcal{V}}

\begin{document}

\title{Computational Complexity of interacting electrons\\ and 
fundamental limitations of Density Functional Theory}

\author{Norbert Schuch}
\affiliation{Max-Planck-Institut f\"ur Quantenoptik,
Hans-Kopfermann-Str.\ 1, D-85748 Garching, Germany.}
\author{Frank Verstraete}
\affiliation{Fakult\"at f\"ur Physik, Universit\"at Wien,
Boltzmanngasse 5, A-1090 Wien, Austria.}

\maketitle

\textbf{One of the  central problems in quantum mechanics is to determine
the ground state properties of a system of electrons interacting via the
Coulomb potential.  Since its introduction by Hohenberg, Kohn, and Sham~%
\cite{DFT-papers-1,DFT-papers-2}, 
Density Functional Theory (DFT) has
become the most widely used and successful method for simulating
systems of interacting electrons, making their original work one of the
most cited in physics. In this letter, we show that the field of
computational complexity imposes fundamental limitations on DFT, as an
efficient description of the associated universal functional would allow
to solve any problem in the class \QMA\ (the quantum version of \NP)
and thus particularly any problem in \NP\ in polynomial time.  This
follows from the fact that finding the ground state energy of the Hubbard
model in an external magnetic field is a hard problem even for a quantum
computer, while given the universal functional it can be computed
efficiently using DFT.  This provides a clear illustration how the field
of quantum computing is useful even if quantum computers would never be
built.}

The difficulty of finding the ground state properties of a large
system of interacting electrons originates both from the exponential
dimension of the underlying Hilbert space and from the fermionic
nature of the wave function. It is a problem encountered virtually everywhere
in quantum chemistry as well as in condensed matter physics: for
instance, the spatial configuration of a molecule is the one for which
the energy of the interacting electrons moving in the nuclear
potential, together with the electrostatic energy of the nuclei,
becomes minimal. Similarly, a rich variety of phenomena in solid
state physics, in particular conductance and magnetic phenomena, can
be understood by considering electrons moving in the periodic
lattice potential, including such exciting phenomena as
high-temperature superconductivity and the fractional quantum Hall
effect.

A system of $N$ electrons is described by the Hamiltonian
\begin{equation}
\label{gen-ham}
H=\underbrace{-\tfrac12\sum_{i=1}^N\Delta_i}_{=:T}+
    \underbrace{\sum_{1\le i<j\le N}
    \frac{\gamma}{|r_i-r_j|}}_{=:I}+\sum_{i}V(x_i)\ 
\end{equation}
($\gamma>0$, and $x_i=(r_i,s_i)$ with $r_i$ position and $s_i$ spin),
where the potential $V$ contains both an electrostatic field $\phi(r)$
and a magnetic field $\vec B(r)$ which couples to the spin (the
coupling to the orbit can be neglected for our purposes, see Supplementary
Material), and the problem is to find the ground state within the set
of fermionic (i.e.\ antisymmetric) quantum states. 
Following the early work of Slater~\cite{DFT-slater}, Hohenberg, Kohn,
and Sham~\cite{DFT-papers-1,DFT-papers-2} showed that this problem could be
rephrased as a single-particle minimization problem,
 for the reason
that the only problem-dependent part is the external potential $V$
whose expectation value only depends on the local density, while the
kinetic and interaction terms $T$ and $I$ are fixed and universal
for all systems. Thus, the ground state energy is given by
\begin{equation}
\label{dft-min}
E_0=\min_{\rho}\left\{\tr(V\rho)+F[\rho]\right\}\ ,
\end{equation}
where $\rho$ is a single-electron density, and the functional $F$
contains the problem-independent minimization over $T$ and $I$,
\begin{equation}
\label{eq:ufunc}
F[\rho]=\min_{\Omega\rightarrow\rho}\tr\left[(T+I)\Omega\right]\ .
\end{equation}
Here, the minimization runs over all $N$-electron density operators
$\Omega$ which give rise to the reduced density $\rho$.  The central
requirement for a good DFT algorithm is to find a suitable approximation
to the universal functional, and indeed better and better techniques have
been developed, making DFT the most widely used and most successful
algorithm for treating interacting electrons.

However, as we show in this letter, there exist fundamental limits
which constrain the ability to find a generally applicable and
efficiently computable approximation to the universal functional,
and thus put bounds on the applicability of DFT.  To this end, we
consider the 2D Hubbard model with local magnetic fields, which arises
from the problem of interacting electrons for a specifically chosen
lattice potential, and can thus be simulated using DFT. We first determine
the computational complexity of solving the Hubbard model and
show that it is among the hardest problems in the complexity class \QMA,
Quantum Merlin Arthur. \QMA\ contains problems which are believed to be
hard to solve even by quantum computers, but once a solution is found, 
it can be checked efficiently by a quantum computer.
Thus, \QMA\ encompasses the complexity class \NP. 
We compare this to the difficulty of solving the Hubbard model
using DFT with a suitable approximation of the functional at hand, and
find that in that case the Hubbard model can be solved by a classical
computer in a time polynomial in the number of electrons.  This means that
the existence of an efficient approximation to the functional would imply
\QMA=\textsf{P}, 
i.e.\ computing the functional to polynomial accuracy in the number of
electrons is a \QMA-hard
problem, which poses
fundamental limitations on the ability to approximate the
functional in DFT.  Of course, this does not mean that DFT is not
applicable in practice: much lower (e.g.\ constant) accuracies will
typically suffice, and DFT is indeed a highly successful method.

The 2D Hubbard model~\cite{hubbard-1,hubbard-2} describes a system of
fermions hopping on  a lattice. Although it typically appears as a
phenomenological model for strongly bound electrons in solid state
physics~\cite{auerbach}, it can be derived rigorously from
(\ref{gen-ham}) for an appropriate potential, as we show in the
Supplementary Material.  
The Hubbard model with local magnetic fields
is given by the Hamiltonian
\begin{equation}
\label{hubb-ham}
H_\mathrm{Hubb}=
    -t\!\!\!\sum_{<i,j>,s}\!\!\!a_{i,s}^\dagger
    a_{j,s}+U\sum_{i}n_{i,\uparrow}n_{i,\downarrow}
    -\sum_{i}\vec\sigma_i\cdot\vec B_i\ ,
\end{equation}
where $a_{i,s}^\dagger$ creates an electron of spin
$s\in\{\uparrow,\downarrow\}$ on lattice site $i$,
\mbox{$<\!i,j\!>$} denotes nearest neighbors on the 2D square
lattice, $n=a^\dagger a$,
$\vec\sigma_i=(\sigma^{x,i},\sigma^{y,i},\sigma^{z,i})$, and 
$\sigma^{\alpha,i}=\sum_{s,s'}\sigma_{ss'}^\alpha a_{i,s}^\dagger a_{i,s'}$
with $\sigma^\alpha$ the Pauli matrices.
 The first term describes an electron
tunneling from one site to the adjacent one without changing its
spin, the second the on-site Coulomb repulsion between two electrons
of different spin sitting on the same site, and the rightmost term
contains the contribution from the magnetic field which imposes a
local field at each site $i$ -- this is the only term which we can
tune locally.

\begin{figure}[t]
\includegraphics[width=3cm]{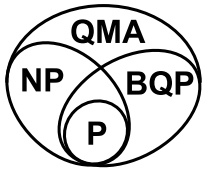}
\caption{The relevant complexity classes and their relations. While
\textsf{P} and \textsf{BQP} are the classes of problems efficiently
solvable by classical and quantum computers, respectively, \NP\ (\QMA)
contains decision problems which are likely to be hard to solve by
classical (quantum) computers, but where for positive instances, 
classical (quantum) proofs exist which can be checked efficiently by a
classical (quantum) computer.  All inclusions are believed to be strict.
We show that solving the Hubbard model is among the hardest problems in
\QMA, while the existence of an efficient description of the universal
functional in DFT would put it in \textsf{P}, leading to the collapse of
all aforementioned complexity classes. This puts tight bounds on the
existence of such functionals.}
\end{figure}

The 2D Hubbard model is of large interest on its own, as it is the
minimal model that is believed to describe the physics arising in
high-temperature superconductivity, quantum magnetism, and heavy
fermions. Indeed, it is one of the most intensively studied models
in solid state physics, making the investigation of its
computational complexity interesting on its own. In the following,
we show that computing its ground state energy up to polynomial
accuracy is complete for the complexity class \QMA, the
quantum analogue of \NP. A decision problem is in \QMA\ if -- although
possibly hard to solve even by a quantum computer -- every positive
instance 
has a \emph{quantum} proof which can be checked efficiently by a quantum
computer. In particular, finding the ground state energy of a local
spin system with an accuracy polynomial in the lattice size
is in \QMA: The ground state
serves as a proof, as expectation values of local Hamiltonians can be
estimated efficiently.  Conversely, it has been shown that any circuit
verifying a \QMA\ proof can be encoded as a ground state
problem~\cite{kitaev-lh-1,kitaev-lh-2}, i.e., ground state problems are
\QMA-complete.  (A problem is called complete for a class if it is among
the hardest problems in this class, i.e., if any problem in the class can
be reduced to it.)  Using the same argument as before, finding the
ground state energy of the Hubbard model is inside \QMA, since it
can be mapped to a spin system via the Jordan-Wigner transform:
This allows to specify its ground state using spins, 
in such a way that it is possible to measure the ground
state energy efficiently~\cite{liu}.

In the following, we show that the Hubbard model with magnetic fields 
is also a hard problem
for \QMA, and thus \QMA-complete.  To this end, we start from a class of
Hamiltonians for which finding the ground state energy is known to be
\QMA-complete -- i.e., as hard as finding the ground state energy of any
local Hamilonian -- and show that this problem can be reduced to
finding the ground state energy of the Hubbard model with local magnetic
fields. This is accomplished by a sequence of reductions, each of which
reduces the previous Hamiltonian problem to a more restricted class of
Hamiltonians. Each step makes use of perturbation theory constructions
(so-called \emph{gadgets}) such that the original Hamiltonian arises as
the effective low-energy theory of the new Hamiltonian.

We start off with the Hamiltonian
\begin{equation}
\label{eq:h_2d}
H_\mathrm{Pauli}=\sum_{<i,j>}\lambda_{ij}A_{(ij)}\otimes B_{(ij)}
\end{equation}
defined on a 2D lattice with $N$ spins, with $A_{(ij)}$ and $B_{(ij)}$
Pauli matrices and $|\lambda|\le1$, for which it has been proven that
finding the ground state energy up to a polynomial accuracy $1/q(N)$ is
\QMA-hard~\cite{oliveira}.  Following Ref.~\cite{oliveira},  we call
interactions of the form $\lambda_{ij}A_{(ij)}\otimes B_{(ij)}$ Pauli
interactions.

We first show how the Pauli Hamiltonian (\ref{eq:h_2d}) can be reduced to
the 2D Heisenberg lattice with local fields (see Supplementary Material
for details).  To this end, we employ a chain of gadgets, all of which
replace a two-qubit coupling by a chain of three qubits with a 
more restricted coupling.  The idea is that by imposing a strong local 
field on the central (``mediator'')
qubit, the system will essentially be in the ground state of the central
qubit -- but there will be second-order processes in which an excitation
hops from the left qubit to the central one and then to the right, or vice
versa, yielding an effective coupling between the outer qubits. (The
excitation can also hop back and give an extra local term, which however
can easily be compensated by adjusting the local magnetic field.) Note
that similar gadgets have already been used, e.g., in
Refs.~\cite{oliveira,biamonte}.

\begin{figure}[t]
\includegraphics[width=.7\columnwidth]{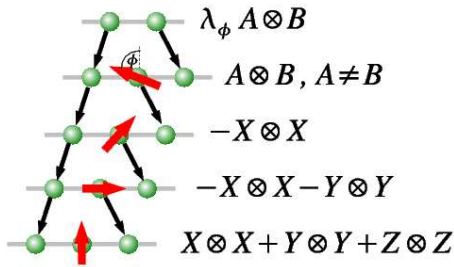}
\caption{
\label{figgadgets}
Gadgets to reduce Pauli couplings to Heisenberg couplings. Each gadget
works by inserting an extra spin in the middle which is subject to a
strong local field, yielding the desired interaction in second order
perturbation theory. $A$ and $B$ are Pauli
matrices.
}
\end{figure}

The full sequence of reductions to the Heisenberg lattice is illustrated
in Fig.~\ref{figgadgets}. In a first step, we reduce arbitrary Pauli
couplings $\lambda A\otimes B$ to Pauli couplings with constant $\lambda$
and $A\ne B$.  We illustrate this with a $\lambda Y_l\otimes Z_r$ coupling (we
use $X$ for $\sigma^x$ etc.\ in the following), which is obtained from
three qubits with couplings $Y_l\otimes X_m\otimes \openone_r+
\openone_l\otimes Y_m\otimes Z_r$ 
by putting a strong field in the $XY$ plane on the central
qubit: A short calculation shows that this indeed gives a $Y_l\otimes Z_r$
coupling, where the strength is given by the angle in the $XY$ plane. The
intuition behind is that e.g.\ a $Y_l$ on the left qubit can excite the
central qubit, and the excitation then hops to the right qubit as a $Z_r$.
In order for this hopping to be possible, the central field must not be
along the $X$ or $Y$ axis, and correspondingly, the hopping amplitude is
controlled by the overlap of the field with $X$ and $Y$ eigenvectors.

The second gadget reduces Pauli couplings $A\otimes B$ ($A\ne B$) to Ising
couplings.  This is achieved by essentially the same gadget as before: For
$X_l\otimes Y_r$, take $X_l\otimes X_m\otimes\openone_r+
\openone_l\otimes Y_m\otimes Y_r$
and place a strong field in $X+Y$ direction on the $m$ qubit.  
In a next gadget,
Ising couplings are reduced to two $X\otimes X+Y\otimes Y$ couplings:
Placing a strong $Y$ field on the central qubits only allows for the
hopping of excitations via the $X\otimes X$ part of the coupling.
Similarly, the above coupling is reduced to the Heisenberg interaction
$X\otimes X+Y\otimes Y+Z\otimes Z$: A strong $Z$ field prohibits hopping
via the $Z\otimes Z$ term, whereas $X\otimes X+Y\otimes Y$ describes
hopping in the $\{\ket{01},\ket{10}\}$ subspace, which to second order
yields the very same hopping term between the two outer qubits. 

Putting these gadgets together, we have managed to reduce the
\QMA-complete Hamiltonian 
(\ref{eq:h_2d}) 
to the Heisenberg Hamiltonian 
in a magnetic field
on a sparse
lattice. This can, in turn, be reduced to the full 2D Heisenberg lattice
by using an ``erasure gadget'': putting strong fields on the qubits to be
erased decouples them up to polynomial precision. We have thus shown that
the 2D Heisenberg Hamiltonian in a magnetic field on a 2D square lattice,
\begin{equation}
\label{heis-ham}
H_\mathrm{Heis}=J\sum_{<i,j>}\vec\sigma_i\cdot\vec\sigma_j-
    \sum_i\vec B_i\cdot\vec\sigma_i\ ,
\end{equation}
is \QMA-complete, both for $J>0$ (which we use further on) and $J<0$.
Note that the presence of a magnetic field is crucial for the
construction, as it is the only set of parameters available to encode a
computational problem.

The final step is to reduce the Heisenberg lattice to the Hubbard model
(\ref{hubb-ham}). The procedure can be found, e.g., in
Ref.~\cite{auerbach}, and has been included in the supplementary material:
In (\ref{hubb-ham}), one chooses an on-site repulsion $U$ very large as
compared to $t$, and operates the system in the so-called half-occupancy
regime where there are as many electrons as sites. (Note that this implies
that a polynomial scaling in the lattice size equals a polynomial scaling
in the number of electrons.) The tunneling is supressed as $t/U$, so that
in the ground state each site will be occupied by exactly one electron,
providing the desired spin degree of freedom. The coupling between the
spins is achieved by a second-order process where one electron tunnels to
an adjacent site, interacts with the other electron, and tunnels back.
However, this can only take place if the spins form a singlet, giving rise
to the effective Hamiltonian (\ref{heis-ham}) up to a constant.  As the
process is of second order, we have that $J=t^2/U>0$, and the error from
higher order processes is $O(N^3t^3/U^2)$; thus, $U/t$ has to grow
polynomially with the system size $N$.

\begin{figure}[t]
\includegraphics[width=0.9\columnwidth]{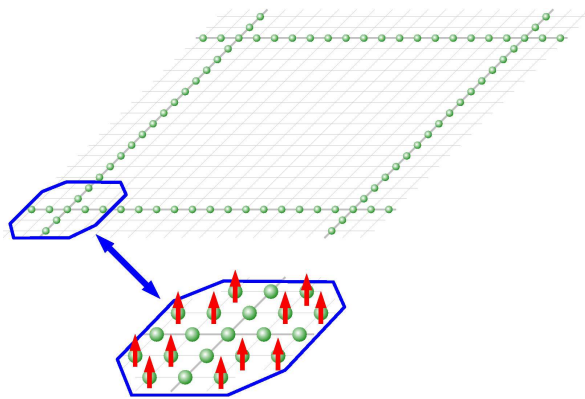}
\caption{The sparse Heisenberg lattice as obtained from
$H_{2D}$, Eq.~(\ref{eq:h_2d}), using a sequence of gadgets.  It can be
reduced to a 2D Heisenberg lattice using the erasure gadget, where strong
local fields are used to decouple unwanted qubits to leading order, as
shown in the inset.
\label{figsparse}}
\end{figure}

All these gadgets can be combined straightforwardly: Firstly, the gadgets
in one layer do not interact, as they never share a coupling term.  It can
thus be checked straightforwardly that to second order perturbation
theory, there will be no cross-talk between the gadgets. Secondly, all
gadget layers can be applied one after another, as long as the
\emph{total} strength of the previous gadgets is sufficiently smaller than
the strong local fields of the new gadgets. Since the number of layers is
constant, this can be achieved by choosing poly-scale field strengths,
which allows for a polynomial scaling of the interaction strength as well
as an arbitrary polynomial precision in energy. (See Supplementary
Materials for details.)

Let us now turn our attention back to DFT and the problem of interacting
electrons. As we show in the Supplementary Material, the Hubbard model
(\ref{hubb-ham}) with arbitrary local fields arises from (\ref{gen-ham})
for an appropriately chosen $V$.  For this particular potential, one can
explicitly write down the wave function $w_i(r)$ of each mode $a_i$ to
sufficient precision.  Thus, the ground state wave function of the Hubbard
model is supported by the $w_i(r)$, and consequently the single-electron
density for the ground state of the Hubbard model must be of the form
$\rho(r)=\sum \lambda_{i,s,s'} |w_i(r)|^2 \ket{s}\bra{s'}$.  Since the
functional (\ref{eq:ufunc}) is convex and the physical $\lambda_{i,s,s'}$
form a convex set in $\mathbb R^{4N}$, the minimization (up to polynomial
accuracy) can be carried out efficiently,\cite{convexopt} 
i.e., finding the ground state energy of the
Hubbard model is in \textsf{P}. This implies that
$\textsf{P}^\textsf{UF}=\QMA$ with $\textsf{UF}$ an oracle for the
universal functional, i.e.\ computing the functional is \QMA-hard under
Turing reductions.

Let us note that there are alternative ways to define $F[\rho]$, e.g.\ as
the minimum over all \emph{pure} $N$-electron
states~\cite{DFT-book-1,DFT-book-2}, in which case $F$ is not convex. Yet,
efficient computability (or even certifiablity) of $F$ would still imply
that one could give a certificate for the ground state energy, i.e.\ \QMA\
would collapse to \NP. This is considered very unlikely, thus implying
that any reasonably defined $F$ cannot be computed in \NP.

Finally, DFT can also be based on a functional defined on two-electron
densities, which can be computed efficiently~\cite{DFT-book-1,DFT-book-2}.
In this case, the \QMA-hardness of the problem arises from the fact that
characterizing the set of allowed two-electron reduced states, the
$N$-representability problem, is \QMA-hard~\cite{liu}; in fact, this
provides an alternative proof of its hardness.

\textbf{Acknowledgements.}
We thank H.\ Buhrman,
G. Burkard,
I.\ Cirac, G.\ Giedke, J.\ Kempe, 
G.\ Refael,
R.\ Schmied,
B.\ Toner, and the referees
for helpful
discussions and comments.
N.S.\ thanks the Erwin Schr\"odinger Institute in Vienna,
where parts of this work were carried out, for their hospitality.

\section*{APPENDIX: NP-completeness of Hartree-Fock}

One of the precursors of DFT which is still widely used 
is the Hartree-Fock method.
 It is similar to DFT in that
it reduces the $N$-electron equation to a problem of individual electrons
moving in an external field which depends only on the average electron
distribution. Different from DFT, Hartree-Fock is based on a
particular ansatz for the wave function and is therefore not guaranteed to
give the true ground state energy; on the other hand, it is an
iterative method which can be applied without prior assumptions, whereas in DFT
some \emph{a priori} guess on the form of the universal functional has to
be made.

The starting point is  a Hamiltonian
\begin{equation}
\label{hf-ham}
\mathcal H=\sum H^{(1)}_{i,j}a^\dagger_ia_j+
    \sum H^{(2)}_{ij,kl}a^\dagger_ia^\dagger_ja_ka_l+\dots\ ,
\end{equation}
which is local, meaning the series terminates at some fixed $H^{(r)}$, and
with a number of fermionic modes $M\ge N$; if it is derived from
two-particle interactions as in the Schr\"odinger equation
(\ref{gen-ham}), $r=2$.  The Hartree-Fock method
approximates the ground state of this Hamiltonian using the ansatz
$b^\dagger_N\cdots b^\dagger_1\ket\Omega$ with
$b_i=\sum u_{ij} a_j$  (where $\ket\Omega$ is the vacuum). Note that this
corresponds to an antisymmetrized product of single-particle wave
functions, which is how Hartree-Fock is usually presented.

In the following, we show that approximating the ground state energy using
the Hartree-Fock method is an \NP-complete problem. More precisely, we
consider the problem of deciding whether the lowest energy of
(\ref{hf-ham}) within the Hartree-Fock ansatz is below some $a$ or above
some $b>a$. We show that the problem is inside \NP\ for up to an 
exponential accuracy $b-a$ and for any $r$, and that for \NP-completeness
a polynomial accuracy $b-a<1/\poly(N)$ and $r=2$ are sufficient.

To see that the problem is in \NP, note that a Hartree-Fock state is
fully characterized by the $u_{ij}$'s, and that from there its energy can
be computed efficiently. Conversely, the problem is shown to be \NP-hard by
mapping it to the ground state problem for Ising spin glasses which is
known to be \NP-hard: Given an $L\times L\times 2$ lattice
of two-level spins $S_i=\pm 1$ with a nearest neighbor
Ising coupling $\mathcal H=\sum J_{ij}S_iS_j$, $J_{ij}\in\{0,-1,1\}$,
determine whether the ground state energy is the minimum one allowed by
the individual $J_{ij}$'s or not. 
Therefore, embed the $N=2L^2$ classical spins into a fermionic
system with $2N$ modes occupied by $N$ fermions. The modes come in pairs
$(a_{2i},a_{2i+1})$, and a Hamiltonian term $\lambda
n_{2i}n_{2i+1}$, $\lambda=O(N^2)$ penalizes double occupancy, so that
in the ground state exactly one mode per pair is occupied, giving an
effective spin degree of freedom~\cite{liu}; the coupling
$J_{ij}S_iS_j$ of these spins is realized as 
$J_{ij}\sum_{p,q=0,1}(-1)^{p+q} n_{2i+p}n_{2j+q}$.
 As the ground state of the system is a classical spin state,
it can be expressed as a Hartree-Fock state where $b_i=a_{2i}$ or
$b_i=a_{2i+1}$, respectively, and since the classical Hamiltonian has a
constant gap while perturbations from the penalized
subspace are at most $O(1/\lambda^2)$, a polynomial
accuracy is sufficient to make the problem \NP-hard.

\vspace{1cm}

\begin{center}
\textbf{SUPPLEMENTARY MATERIAL}
\end{center}

\vspace{0.6cm}

\noindent\textbf{1.\ Second order perturbation theory}

\vspace{0.3cm}

We start with a Hamiltonian $H=H_0\oplus H_1$ and a perturbation 
\[
V=\left(\begin{array}{cc}V_0&V_{01}\\V_{10}&V_1\end{array}\right)
\]
with $\|H_0\|,\|V\|\le v$ and $H_1\ge\Delta\gg v$,
and want to show that
the low-energy spectrum of $H_\mathrm{tot}=H+V$ is well approximated by 
$H_\mathrm{eff}=H_0+V_0-V_{01}H_1^{-1}V_{10}$. To this end,
rotate $H_\mathrm{tot}$ by a unitary $U=e^{S}$, 
\[
S=\left(\begin{array}{cc}0&X\\
    -X^\dagger&0\end{array}\right)\ ,
\]
where $X=-V_{01}H_1^{-1}+V_{01}H_1^{-1}V_1H_1^{-1}-
    V_0V_{01}H_1^{-2}-H_0V_{01}H_1^{-2}$ is
chosen such as to make the Hamiltonian as diagonal as
possible. 
For systematic constructions of $S$ for any 
order of perturbation theory, see Ref.~\cite{schrieffer-wolf}.
By expanding $S$ to second order in $v/\Delta$, we obtain a new
Hamiltonian 
\[
\tilde H_\mathrm{tot}=
UH_\mathrm{tot}U^\dagger=
\left(\begin{array}{cc}H_\mathrm{eff}+O(\tfrac{v^3}{\Delta^2})
    & O(\tfrac{v^3}{\Delta^2}) \\
    O(\tfrac{v^3}{\Delta^2}) & H_1+V_1+O(\tfrac{v^2}\Delta)
\end{array}\right),
\]
where all $O(\cdot)$ symbols are bounds in operator norm.
Now compare this Hamiltonian with the block-diagonal Hamiltonian 
$\tilde H_\mathrm{diag}$ obtained from
$\tilde H_\mathrm{tot}$ by setting the off-diagonal blocks to zero: The
low-energy spectrum of $\tilde H_\mathrm{diag}$ is given by
$H_\mathrm{eff}+O(\tfrac{v^3}{\Delta^2})$, and since
$\|\tilde H_\mathrm{diag}-\tilde
H_\mathrm{tot}\|_\mathrm{op}=O(\tfrac{v^3}{\Delta^2})$,
it follows that all eigenvalues are $O(\tfrac{v^3}{\Delta^2})$\,-\,close to
each other, and thus the low-energy spectrum of $H+V$ is given by 
\begin{equation}
\label{eq:2nd-order-pert}
H_\mathrm{eff}=H_0+V_0-V_{01} H_1^{-1} V_{10}+O(\tfrac{v^3}{\Delta^2})\ .
\end{equation}
Note that when applying the gadgets to an $N$-qubit system with an
extensive number of local perturbations, the error bound will depend on
$N$ since $v\propto N$.

\vspace{0.6cm}

\noindent\textbf{2.\ Gadgets}

\vspace{0.3cm}

Let us now turn towards the gadget constructions.
Starting from (\ref{eq:h_2d}), we want to show how its low-energy
subsector can be
obtained as an effective theory from (\ref{hubb-ham}). Let us first note
that all perturbation gadgets of one level can be applied simultaneously,
since they are second order gadgets and the transition term $V_{10}$ is a
sum of two-body terms which excite \emph{one} qubit only; to return to the
ground state subspace in the next step, this excitation has to hop to one
of the adjacent sites. Thus, 
there will be no cross-gadget terms and
the action of each layer of gadgets can be
investigated on the level of a single gadget.

We aim to approximate (\ref{eq:h_2d}) with strength $\lambda_{ij}\le 1$ up
to a precision $O(1/q)$, with $q\equiv q(N)$ a polynomial in $N$ (we will in
the following omit the parameter $N$ for most polynomials).  For $A$ and
$B$ Pauli matrices, we can obtain a tunable Pauli coupling
$\lambda_T A\otimes B$ from 
\begin{equation}
V=\lambda_{P}A\otimes
X\otimes\openone +\lambda_P\openone\otimes Y\otimes B\ ,
\label{eq:diff-pos-pauli-cpl}
\end{equation}
by acting with a Hamiltonian $H=B_P\ket{e_\phi}\bra{e_\phi}$,
$B_P\gg\lambda_P$, and the excited state
$\ket{e_\phi}=\ket0-e^{i\phi}\ket1$.  Then, to second order, the system is
described by the effective Hamiltonian
\[
H_T=+2\lambda_P^2/B_P\sin\phi\cos\phi\; A\otimes B+O(\lambda_P^3/B_P^2)
\]
(up to a constant, and times the ground state projector on the middle
qubit); note that when combining the gadgets, the total error grows with
the third power of the total strength of $V$, and thus as $N^3$. As we aim
to implement any $|\lambda_T|\le1$, we set $\lambda_P^2=B_P$ and tune the
actual value using $\phi$.  Choosing
$\lambda_P=N^4q$, $B_P=N^8q^2$ [$N^4q\equiv N^4q(N)$], we find that the
\emph{total} error is at most $N^3O(\lambda_P^3/B_P^2)=O(1/Nq)$ and thus much
below the targeted precision $O(1/q)$.  Note that in particular, this
allows to split any Pauli interaction in two interactions of the form
$X\otimes Y$, i.e.\ with two different Pauli matrices and positive sign.

Let us now show how such an $X\otimes Y$ coupling as in
Eq.~(\ref{eq:diff-pos-pauli-cpl}) can be reduced to Ising interactions. To
this end, consider the Hamiltonian 
\begin{equation}
\label{eq:ising-cpl-ham}
V=-\lambda_IX\otimes X\otimes \openone-\lambda_I\openone\otimes Y\otimes Y
+H_\mathrm{loc}
\end{equation}
and apply a field 
$H=B_I\ket{e_{\pi/4}}\bra{e_{\pi/4}}$.
Here, $H_\mathrm{loc}$ represents the local fields of the preceding gadget
layers. They act on the qubits remaining after the present gadget, i.e.\
do not induce transitions to excited states, and are thus 
first-order terms 
which are left untouched in (\ref{eq:2nd-order-pert}). We choose 
$B_I=N^{20}q^5\gg N\lambda_I$, and 
$\lambda_I=N^{12}q^3\gg \|H_\mathrm{loc}\|= B_P=N^9q^2$,
which results in an effective Hamiltonian
\[
H_P=+\lambda_I^2/B_I X\otimes Y+O( \lambda_I^3/B_I^2)+H_\mathrm{loc}\ ,
\]
for which $\lambda_I^2/B_I=\lambda_P$, and $N^3O(\lambda_I^3/B_I^2)=1/Nq\ll
O(1/q)$. Note that due to rotational invariance, this construction holds
for any type of Pauli coupling.

Ising interactions Eq.~(\ref{eq:ising-cpl-ham}) can in turn be reduced to
$XX$-type interactions, 
\begin{equation}
\label{eq:XX-cpl-ham}
V=-\lambda_{XX}(X\otimes X+Y\otimes Y)\otimes\openone-
    \openone\otimes(X\otimes X+Y\otimes Y)+H_\mathrm{loc}
\end{equation}
by putting a field in the $Y$ direction,
$H=B_{XX}(\openone-Y)/2$. This cancels all $Y$ contributions in the
Hamiltonian, since $\bra{0_y}Y\ket{1_y}=0$, and one remains with the $X$
part of $V$,
\[
H_{XX}=-2\lambda_{XX}^2/B_{XX}X\otimes
X+H_\mathrm{loc}+O(\lambda_{XX}^3/B_{XX}^2)\ .
\]
(The factor $2$ is due to the fact that either an $X$ on the left can
excite the middle qubit, which then decays towards the right, or vice
versa.) We choose $\lambda_{XX}=N^{28}q^7/4$ and $B=N^{44}q^{11}/8$, which
ensures that $2\lambda_{XX}^2/B_{XX}=\lambda_I$, $B_{XX}\gg\lambda_{XX}\gg
B_I$, and the total error is $1/Nq\ll O(1/q)$, as required.

In a last step, we reduce the Hamiltonian with $XX$ type couplings
to an antiferromagnetic  Heisenberg Hamiltonian with local fields.
To this end, consider 
\begin{align}
V=&\lambda_H\sum_{S=X,Y,Z}
    (S\otimes S\otimes\openone+
	\openone\otimes S\otimes S)+H_\mathrm{loc}-\dots\nonumber\\
&-\lambda_H^2/B_H
(Z\otimes\openone\otimes\openone+\openone\otimes
\openone\otimes Z)
\label{eq:v-for-xx-from-heis}
\end{align}
and place a strong field in $Z$ direction, $H=B_H(\openone-\sigma_Z)/2$,
on the central qubit. Intuitively, the $X\otimes X+Y\otimes Y$ part
describes the hopping of an excitation from one side through the central
qubit to the other side; since the excitation can also hop back to the
original site, it however also induces an additional local field which is
compensated by the extra term in Eq.~(\ref{eq:v-for-xx-from-heis}).
The effective Hamiltonian obtained is then
\[
H_H=-2\lambda_H^2/B_H(X\otimes X+Y\otimes Y)\ ,
\]
and by choosing $B_H=N^{92}q^{23}/512$, $\lambda_H=N^{60}q^{15}/64$, we find
that $2\lambda_H^2/B_H=\lambda_{XX}$, $B_H\gg\lambda_H\gg B_{XX}$, and the
total error is again $O(1/Nq)$.

By combining these gadgets, we find that each Pauli coupling can be
reduced to a line of $16$ Heisenberg couplings with variable local fields.
Note that it should be possible to significantly reduce the order of
magnitude of the fields by going to higher order perturbation theory: Each
second order gadget couples the two outer qubits by an excitation hopping
through the middle qubit. Therefore, it should be possible to choose all
fields of equal magnitude and go to $16$th order perturbation theory,
which is the lowest non-vanishing order, and 
to which solely hopping terms contribute.
 Note further that one can decrease the length of the
chain to $12$ couplings (and thus to $12$th order perturbation theory), as
one can equally combine one $XY$ Pauli and one Ising interaction to obtain
an arbitrary Pauli coupling, including antiferromagnetic Ising couplings.

\vspace{0.6cm}

\noindent\textbf{3.\ Erasure gadget}

\vspace{0.3cm}

The sparse Heisenberg lattice
Fig.~\ref{figsparse}a can
be straightforwardly reduced to a full 2D Heisenberg lattice with local
fields. To this end, add fields $H=B_e(1-\sigma_z)/2$ on
all qubits to be erased, while
\[
V=\lambda_H\sum_{<ij>}\vec\sigma_i\cdot\vec\sigma_j +\sum_i\vec
B_i\cdot\vec\sigma_i
\]
is the 2D Heisenberg lattice. Then, 
according to Eq.~(\ref{eq:2nd-order-pert}), 
$H_\mathrm{eff}=H_0+V_0+O(\|V\|^2/B_e)$, which yields the Heisenberg
Hamiltonian on the sparse lattice. In particular, given that 
$\|V\|\le N\lambda_H$, by choosing $B_e=N^3\lambda_H^2q$ we find that 
$B_e\gg V$, and the total error is $O(1/Nq)$.

\vspace{0.6cm}

\noindent\textbf{4.\ Reduction from Heisenberg to Hubbard model}

\vspace{0.3cm}

The final reduction
step shows how the Heisenberg model can be reduced to the Hubbard
model Eq.~(\ref{dft-min}) (see, e.g.,
Ref.~\onlinecite{auerbach}). 
To this end, choose an one-dimensional ordering of
the Hubbard lattice, e.g.\ row-wise from left to right, and always place
the spin-up mode before the spin-down mode. This results in a
one-dimensional ordering of the modes of the Hubbard model,
$(a_{1,\uparrow},a_{1,\downarrow},a_{2,\uparrow},a_{2,\downarrow},\dots)$.
Now apply a Jordan-Wigner transform, 
\[
a_{i,s}\rightarrow\left(
    \prod \sigma^z_{i',s'}\right) \sigma^{-}_{i,s}
\]
where the product runs over all $(i',s')$ left of $(i,s)$. 
This transforms (\ref{hubb-ham}) to a two-level system with a Hamiltonian
$H+V$,
\begin{align*}
H=&U\sum_{i}n_{i,\uparrow}n_{i,\downarrow}\\
V=&-t\sum_{<i,j>,s}\sigma_{i,s}^{+}\big[\Pi\,\sigma_{k,s'}^z\big]\sigma_{j,s}^-\\
&+\sum_j\left[(B^x_j-iB^y_j)\sigma_{j,\uparrow}^+
\sigma_{j,\downarrow}^-+\mathrm{h.c.}\right]
+ B_z(n_{j,\uparrow}-n_{j,\downarrow})
\end{align*}
with $n=\sigma^+\sigma^-=\ket1\bra1$. We consider $V$ as a perturbation to
$H$, i.e.\ $U\gg t,\vec B$, and do a second-order expansion. Since we
operate the system in the half-occupancy regime, the ground state of $H$
satisfies $n_{i,\uparrow}+n_{i,\downarrow}=1$, which makes the $\sigma^z$
string in the tunneling term vanish on all but the sites $i$ and $j$.
The half-occupancy allows to interpret the ground-state subspace as a
system of spin $\tfrac12$ particles by grouping modes $(i,\uparrow)$ and
$(i,\downarrow)$.  The magnetic term in $V$ contributes only to first
order (and yields the magnetic field operator on the resulting two-level
system), so that the second-order term is found by considering four
sites $((l,\uparrow),(l,\downarrow),(r,\uparrow),(r,\downarrow))$, with a
Hamiltonian $H+V$, 
\begin{align*}
H=&U(\ket{11}\bra{11}_l\otimes\openone_r+\openone_l\otimes\ket{11}\bra{11}_r)\
, \\
V=&-t(\sigma^+\otimes\sigma^z\otimes\sigma^-\otimes\openone+
  \openone\otimes\sigma^+\otimes\sigma^z\otimes\sigma^-+\mathrm{h.c.})\ .
\end{align*}
A straightforward calculation on the subspace $\{\ket{1001},\ket{0110}\}$
-- the only one with non-vanishing second order contributions --
shows that this leads to a term
\[
H=-\frac{4t^2}{U}(\ket{01}-\ket{10})(\bra{01}-\bra{10})
\]
expressed in the effective spin $\tfrac12$'s described above, 
which up to a constant equals the antiferromagnetic Heisenberg Hamiltonian
$(2t^2/U) \vec\sigma\cdot\vec\sigma$.
Choosing $U=N^8\lambda_H^3q^2/8$ and $t=N^4\lambda_H^2q/4$, we have that
$U\gg t,\|B\|=O(NB_e)$, $2t^2/U=\lambda_H$, and the error $N^3t^3/U^2\ll
1/Nq$ as desired.

Let us note that as with the gadgets before, no cross-terms appear when
applying the gadgets together, as the only way to return to the ground
state subspace in second order are processes within a single gadget.

\vspace{0.6cm}

\noindent\textbf{5.\
Reduction of the Hubbard model to the Schr\"odinger equation}

\vspace{0.3cm}

In the following, we show how finding the ground state energy of the
Hubbard model with a local field up to $1/\poly(N)$ precision can be
reduced to answering the same question for the Schr\"odinger equation
(where $N$ is both the number of sites of the Hubbard model and the
number of electrons).

\emph{Structure of the proof.---}%
Let us first give an overview of the proof, highlighting the crucial
steps. In the first part, we use the kinetic term together with an
appropriate external electrostatic potential in the Schr\"odinger equation
[terms $T$ and $V$ in (\ref{gen-ham})] to construct an exactly solvable
model with
the following property: The Hamiltonian can be decomposed as
\begin{equation}
\label{eq:sm:HisH0plusH1}
H=H_0 \oplus H_1\ ,
\end{equation}
where
\begin{equation}
\label{eq:sm:H0H1Ham}
H_0=-t\sum_{\langle i,j\rangle,s} a_{i,s}^\dagger a_{j,s}
+ O(N^{-2\tau+1})
+\mathrm{const.}
\end{equation}
and the constant is chosen such that $H_0\le -\Delta$, and 
with $H_1\ge0$. Note that beyond the gap above the $H_0$ band, we do not
care about the properties of $H_1$.  

In the second part of the proof, we show how to incorporate the magnetic
field and the Coulomb interaction which will yield the on-site repulsion
term.  Loosely speaking, we will treat the Coulomb interaction as a
perturbation to the original Hamiltonian, 
and obtain the on-site repulsion in first order
perturbation theory. However, this cannot be done using the tools for
perturbative expansions used for spin systems (cf.~Sec.~1 of the
Supplementary Material) due to the unbounded nature
of the Coulomb interaction. Instead, we will use a direct estimate to bound
the effect on the ground state energy which stems from off-diagonal
elements of the Coulomb interaction (i.e.\ those coupling the $H_0$ and
the $H_1$ subspace). We then find that the ground state energy of the
Hubbard model with local magnetic fields equals 
the ground state energy of the Schr\"odinger equation with an
appropriately chosen external potential
up to $1/\poly(N)$, as claimed. (Note that the result
obtained by some perturbation expansions is stronger since the whole
low-energy spectrum is reproduced; however, this is not necessary for the
reduction.)

Before we start with the derivation, let us fix the desired scaling of the
variables: We aim to obtain a Hubbard model (\ref{hubb-ham}) with
arbitrary local fields on an $N:=L_x\times L_y$ lattice, and with the
following scaling of the parameters: $t=N^{-\tau}$, $B_\mathrm{max}=\max
|B_i|=O(N^{-\tau})$, $U=\mathrm{const.}\times N^{-\zeta}$, 
and a precision in energy of $O(N^{-2\zeta+2})$, where we have that
$0<\zeta<\tau-3$.  Note that the \emph{relative} accuracy increases as
$\zeta$ and $\tau$ are scaled up, which allows us to obtain the polynomial
accuracies needed for the perturbation gadgets discussed above.

\emph{The exactly solvable hopping model.---}%
We start by constructing the 2D hopping model. We first consider a 1D
exactly solvable model, the Kronig-Penney model, from which we then
construct an exactly solvable model in 3D.
(We set up a 3D lattice
since we consider the Schr\"odinger equation in three-dimensional space;
the same reduction would also work in 2D right away.)
The 1D Kronig-Penney model on $[0,L]$ with periodic boundary conditions
is defined by 
\begin{equation}
\label{eq:sm:V1Ddef}
V(r)=-\V\sum_{n=0}^{L-1}\delta(r-n)\ ,
\end{equation}
where we choose $\V=\tau\log N$.  This model is exactly
solvable: The eigenfunctions are Bloch waves
\begin{equation}
\label{eq:sm:blochwaves}
\psi_k(r+n)=\tfrac{1}{\mathcal N}e^{ikn}
    \left[e^{-\kappa r}+Ye^{-\kappa(d-r)}\right]
\end{equation}
(where $r\in[0,1]$, $n=0,\dots,L-1$),
with
\[
Y=\frac{e^{ik+\kappa}-1}{e^\kappa-e^{ik}}=e^{ik}+O(e^{-\kappa})
\]
and
normalization $\mathcal N^2=L/\V+O(e^{-\V})$.

The dispersion relation for the lowest Bloch band of the Kronig-Penney
model can be approximately solved as
\[
E_k=-\kappa^2=-\V^2-4 \V e^{-\V}\cos(k)+O(\V^2 e^{-2\V})\ .
\]
This band is the only one with bound states, 
with a gap 
of $\Delta=\V^2-O(\V N^{-\tau})$ above.
Expressing the Hamiltonian in the lowest Bloch band 
in terms of the creation/annihilation operators $a_l$ corresponding to 
the Wannier functions $w_l=\sum e^{i k l}\psi_k/\sqrt{L}$,
one finds 
\begin{equation}
\label{eq:hopping-1d}
H_0^{\mathrm{1D}}=-\V^2\sum_l{a_l^\dagger a_l}-
t\sum_{\langle i,j\rangle} a_i^\dagger a_j+O(L\V^2N^{-2\tau})\ ,
\end{equation}
where $t\equiv e^{-\V}=N^{-\tau}$. 

In order to obtain a three-dimensional solvable model, we use a potential
$V(r_1,r_2,r_3)=V(r_1)+V(r_2)+V(r_3)$ with the one-dimensional potentials
of Eq.~(\ref{eq:sm:V1Ddef}).  This choice of the potential leads to a
product ansatz for the wavefunction, where the behavior of the lowest band
is still described by the hopping Hamiltonian (\ref{eq:hopping-1d}), but on a
three-dimensional lattice; the energy gap to the next band is still given
by $\Delta$, the gap of the 1D model.  Using this potential, we can set up a
$L_x\times L_y\times 1$ lattice, $N:=L_xL_y$. 
 Using (\ref{eq:sm:blochwaves}), 
we find that the Wannier functions of the model are of the form 
\begin{equation}
\label{eq:sm:wannierfunc}
w_0(r)=\V^{3/2}e^{-\V |r|_1}+O(\sqrt\V e^{-\V})\ ,
\end{equation}
and $w_i(r)=w_0(r-i)$, where $i=(i_1,i_2)\in\{0,\dots,L_x-1\}
\times \{0,\dots,L_y-1\}$ 
is the site index in the 2D lattice. Thus, we obtain the system
described by (\ref{eq:sm:H0H1Ham}).

Clearly, we can include the spin degree of freedom without affecting the
model at the current stage, as the Hamiltonian currently does not include
any magnetic field. As a result, the
Wannier functions get an an additional spin index,
$w_{n,s}(r)\equiv w_n(r)\otimes\ket{s}$. 

\emph{Treating magnetic field and Coulomb repulsion.---}%
Let us now show how to account for the effect of the magnetic field and
the Coulomb repulsion.  We obtain the magnetic field of the Hubbard model
by putting a magnetic potential 
\[
V_\mathrm{mag}(r)=\sum \vec B_n \chi(r+n)
\]
in (\ref{gen-ham}). Here, $\chi(r)=(1-\exp(-\V))^3$ for $-\tfrac{1}{2}\le 
r_i\le\tfrac{1}{2}$ and zero otherwise. This choice ensures the
following:
\\
i) $\bra{w_{n,s}}V_\mathrm{mag}(r)\ket{w_{n,s'}}= \bra{s}\vec
B_n\cdot\vec\sigma\ket{s'}$ yields the effect of the field $B_n$ on the
spin degree of freedom. 
\\
ii) $\bra{w_{n,s}}V_\mathrm{mag}(r)\ket{w_{m,s}}=O(\V N^{-2\tau+1})$ is
sufficiently small for $n\ne m$, using 
(\ref{eq:sm:wannierfunc}); the unwanted contribution from
the magnetic field is any state is thus $O(\V N^{-2\tau+3})$.
\\
iii) For any state $\ket\chi$,
 $\bra{\chi}V_\mathrm{mag}\ket{\chi}\ge - N^2 B_{\max} =
O(N^{-\tau+2})$.
(This bound can e.g.\ be obtained by neglecting the antisymmetry of
the wave function.)

Before incorporating the Coulomb term $I$, note that the strength $\gamma$
of the Coulomb interaction can be tuned relative to the other terms by
rescaling the spatial coordinates of the system; we choose
$\gamma=N^{-\zeta}/2\V$.  The Coulomb term $I$ has properties very
analogous to those of $V_\mathrm{mag}$:
\\
i) The on-site repulsion is
$\bra{w_{n,0}\otimes w_{n,1}}I\ket{w_{n,0}\otimes w_{n,1}}=
0.8984(\dots)N^{-\zeta}+O(\V^4N^{-\tau-\zeta+2})$
[we explain the calculation of the integral, including the evaluation of
the prefactor, later; the error term is from (\ref{eq:sm:wannierfunc})];
again, by neglecting antisymmetry, this yields a bound $O(N^{-\zeta+2})$
for the total on-site repulsion in the lowest band.
\\
ii) 
$\bra{w_{n,s}\otimes w_{n',s'}}I\ket{w_{m,t}\otimes w_{m',t'}}=
O(\V^4N^{-\tau-\zeta+2})$ 
unless $n=n'=m=m'$, using (\ref{eq:sm:wannierfunc}); the unwanted
cross-terms from the Coulomb repulsion are thus 
$O(\V^4N^{-\tau-\zeta+4})$.
\\
iii) For any state $\ket\chi$ (in particular for any state in the excited
band), $\bra{\chi}I\ket{\chi}\ge 0$.

\emph{Dealing with unbounded perturbations.---}By using the properties
i)-iii) above, we will now be able to show that the ground state energy of
the total Hamiltonian $H_\mathrm{tot}=H_\mathrm{ex}+V_\mathrm{mag}+I$ is
well approximated by the energy of $\Pi_0 H_\mathrm{tot}\Pi_0$, where
$\Pi_0$ projects onto the lowest band of $H_\mathrm{ex}$ (which gives the
first order perturbation expansion).

Let $\ket\psi\equiv\sqrt{1-p}\ket\phi+\sqrt{p}\ket\chi$ be a ground state
of $H_\mathrm{tot}$, where $\ket\phi$ is supported in $\Pi_0$ and
$\ket\chi$ in the orthogonal subspace $\Pi_1=1-\Pi_0$ of high-energy states. We
claim that then, $p$ is very small and thus $\bra\phi
H_\mathrm{tot}\ket\phi$ has almost the same ground state energy, i.e., the
ground state energy of $\Pi_0 H_\mathrm{tot}\Pi_0$ is a good approximation
to the ground state energy of $H_\mathrm{tot}$. (Since we will find that
$p$ is very small, this actually also implies that the ground state of
the projected Hamiltonian is close to the true ground state.)

The error made in the energy by replacing $\ket\psi$ by $\ket\phi$ is
\begin{align}
\nonumber
\Delta E &= \bra\phi H_\mathrm{ex} + V_\mathrm{mag} + I \ket\phi
    -\bra\psi H_\mathrm{ex} + V_\mathrm{mag} + I \ket\psi 
	    \\ \nonumber
&=p\big[\bra\phi H_\mathrm{ex}+V_\mathrm{mag}+I\ket\phi-
    \bra\chi H_\mathrm{ex}+V_\mathrm{mag}+I\ket\chi\big]
	    \\ \nonumber
    &\qquad +2\sqrt{(1-p)p}
    \big[\mathrm{Re}\bra\phi V_\mathrm{mag}\ket\chi
    +\mathrm{Re}\bra\phi I\ket\chi\big]
\end{align}
To bound $\Delta E$, we use the following facts (obtained by combining the
statements about $V_\mathrm{mag}$ and $I$ made before):\\
i) $\bra{\phi}H_\mathrm{ex}+V_\mathrm{mag}+
I\ket{\phi}\le-\Delta+O(N^{-\zeta+2})$;
\\
ii) $\bra{\chi}H_\mathrm{ex}+
    V_\mathrm{mag}+I\ket{\chi}\ge -O(N^{-\tau+2})$;
\\
iii) From the Cauchy-Schwarz inequality,
\begin{align*}
\mathrm{Re}\bra\phi M\ket\chi 
    & \le |\bra\phi M\ket\chi| 
    \\
    &\le \sqrt{\bra\phi M M^\dagger \ket\phi \bra\chi\chi\rangle }
     = \sqrt{\bra\phi M M^\dagger \ket\phi}
\end{align*}
Combining i)-iii), 
 this yields a bound
\[
\Delta E\le p(- \Delta+\alpha) +2\sqrt{(1-p)p}\,\beta
\]
with $\alpha=O(N^{-\zeta+2})$ [from i) and ii)], and 
$\beta^2=\bra\phi I^2+V_\mathrm{mag}^2\ket\phi=
O(N^{-2\zeta+2})$ (the dominating $I^2$ term can be derived solely
from scaling arguments, see later).
Using $\Delta\gg\alpha,\beta$, it is straighforward to show that the
maximum of the above expression [found at $p=O(\beta^2/\Delta^2)$] is
$O(\beta^2/\Delta)=O(N^{-2\zeta+2})$, which bounds the error in the ground
state energy we make by replacing
$H_\mathrm{tot}$ by $\Pi_0 H_\mathrm{tot} \Pi_0$.

\emph{Evaluation of Coulomb energies.---}%
Let us now show how to compute the strength of the on-site repulsion from
the Coulomb interaction. Following (\ref{eq:sm:wannierfunc}), we have to
evaluate the integral 
\[
\V^6\gamma \int \dd^3r \dd^3s \frac{e^{-2\V(|r|_1+|s|_1)}}{|r-s|_2}
= \frac{\V\gamma}{32} 
\underbrace{\int \dd^3r \dd^3s \frac{e^{-(|r|_1+|s|_1)}}{|r-s|_2}}_{c_U}
\]
The latter integral is a constant, $c_U=28.7496(\ldots)$.  Moreover, it is
possible to compute $c_U$ to any accuracy $\epsilon$ in a time
$1/\poly(\epsilon)$ which is sufficient to obtain an efficient reduction.
To this end, first rewrite the integral as
\begin{equation}
\label{eq:sm:greens-int}
c_U=\int \dd^3 q \frac{1}{|q|_2} G(q)\ ,
\end{equation}
where $G(q)$ is the Greens function
\begin{align*}
G(q)&=\int \dd^3 r \dd^3 s\; e^{-(|r|_1+|s|_1)}\delta(r-s-q)
	\\
    & = \prod_i(1+|q_i|)\,e^{-|q_i|}
\end{align*}
Rewriting (\ref{eq:sm:greens-int}) in
spherical coordinates and integrating over $r$, we are left with
\[
c_U=8 \int_0^{\pi/2}\!\!\!\!\!\dd\phi
    \int_0^{\pi/2}\!\!\!\!\!\dd\theta\,\,
    \frac{n(\phi,\theta)}{d(\phi,\theta)}
\]
where $d(\phi,\theta)=((\cos\theta+\sin\theta)\sin\phi+\cos\phi)^5\ge1$
and $n(\phi,\theta)$ are trigonometric polynomials. Since the integrand
and its derivatives are bounded, the integral can be evaluated numerically
to precision $\epsilon$ using a grid of size $1/\poly(\epsilon)$.

\emph{The effective Hamiltonian.---}%
Putting all steps together, we obtain the Hubbard model (\ref{hubb-ham})
with tunneling $t=N^{-\tau}$, on-site repulsion
$U=0.8984(\dots)\,N^{-\zeta}$, and the desired magnetic fields. Collecting
all error terms, one finds that the total error in the ground state energy
is given by $O(N^{-2\zeta+2})$, as desired.

\emph{Remarks.---}A few notes: First, the fact that we are using a
$\delta$-potential for our model does not affect our claims about DFT,
since only the electron density, which is free of singularities, is passed
to the functional; particularly, all these densities arise from
$N$-electron states.  Second, in the Schr\"odinger equation
(\ref{gen-ham}) we have omitted the coupling of the magnetic field to the
orbit of the electrons: the variant of DFT arising from this approximation
is known as ``spin-density functional theory''~\cite{DFT-book-1,DFT-book-2}, 
and our
hardness result holds for exactly this variant. Note also that a coupling
to the orbit would result in a so-called Peierls phase
$e^{i\phi_{kl}}a_k^\dagger a_l$ in the tunneling term of the Hubbard
model, which gives non-vanishing terms only for non-trivial loops, i.e.\
only from fourth order perturbation theory on, and can therefore be
neglected.

\end{document}